\documentclass[amsmath, amssymb, aps, prx, twocolumn, notitlepage]{revtex4-2}
\usepackage[utf8]{inputenc}
\usepackage{graphicx}               % Include figure files
\usepackage[caption=false]{subfig}  % Subfigures
\usepackage{dcolumn}                % Align table columns on decimal point
\usepackage{bm}                     % Bold math symbols
\usepackage[dvipsnames]{xcolor}     % Enable dvips names for color
\usepackage{slashed}                % Feynman slash notation
\usepackage{hyperref}               % Hyperlinked references
\usepackage{enumitem}               % Customizable lists
\usepackage{mathtools}              % Math beautification
\usepackage{floatrow}

\newcounter{myparagraphs}
\begin{document}
\title{Resolving the Corbino Shockley-Ramo Paradox for Hydrodynamic Current Noise}
\author{Aaron Hui}
\affiliation{Department of Physics, Ohio State University, Columbus, Ohio 43202, USA}
\date{\today}

\begin{abstract}
    Johnson noise thermometry enables direct measurement of the electron temperature, a valuable probe of many-body systems. Practical use of this technique calls for non-equilibrium generalizations of the Johnson-Nyquist theorem. For a hydrodynamic Corbino device, however, a na\"ive use of the Shockley-Ramo theorem alongside the ``Corbino paradox'' leads to yet another paradox: current noise through the contacts would seem to be completely insensitive to bulk fluctuations. In this work, we resolve the unphysical ``Corbino Shockley-Ramo paradox'' by correctly formulating the hydrodynamic Shockley-Ramo problem. This allows us to properly formulate the problem of current noise in a hydrodynamic multiterminal device of arbitrary geometry, as well as validate a previously unjustified assumption for rectangular geometry results. As an example, we compute the Johnson noise in a hydrodynamic Corbino device, where we find a suppression of Johnson noise with magnetic field. This unusual characteristic serves as a qualitative signature of viscous hydrodynamic behavior. 
\end{abstract}

\maketitle

\section{Introduction}

The characterization of thermal properties is a venerated tool in probing many-body systems in condensed matter and materials physics. For instance, a sample's thermal conductivity and specific heat are reflective of the nature of its quasiparticles, including those that are charge-neutral. However, accurate metrology of many modern devices requires accurate thermometry at the nanoscale. Specifically for electronic systems, one particular challenge is to disentangle phonon from electron contributions to heat transport; with weak electron-phonon coupling, the electron and phonon temperatures might not even be equal. 

One solution is Johnson noise thermometry: it allows direct and isolated access to the electronic degrees of freedom. The simplest incarnation of Johnson noise thermometry is a direct application of the Johnson-Nyquist theorem (fluctuation-dissipation theorem). For a two-terminal device with resistance $R$ held at uniform electronic temperature $T_0$, it says
\begin{align}
    \lim_{\omega \rightarrow 0} \frac{R}{2k_B} \langle \delta I(\omega) \delta I(t=0) \rangle \equiv T_\text{JN} = T_0
    \label{eq: JN theorem}
\end{align}
where $\delta I(t)$ is the charge current fluctuation at time $t$, $\delta I(\omega)$ is its two-sided Fourier transform \footnote{It is also common to express the Johnson-Nyquist theorem with one-sided power spectral density (i.e. a one-sided Fourier transform), in which case the prefactor changes to $R/(4k_B)$.}, and $\langle ... \rangle$ denotes an ensemble average (time-average if ergodicity is assumed). That is to say, a noise measurement of the Johnson noise temperature $T_{JN}$ is equivalent to measuring the equilibrium electron temperature $T_0$. Furthermore, this thermometry technique requires no calibration; Johnson noise thermometry acts a primary thermometer. This technique has been used to make ultra-sensitive bolometers \cite{Karasik2014, Efetov2018, Miao2018, Liu2018, Miao2021} and to measure of thermal conductivity and heat capacity \cite{Fong2012,Crossno2015,Waissman2022}.

In many practical situations, including the applications above, Eq.~\eqref{eq: JN theorem} does not apply since the electronic temperature is not spatially uniform. Generalizations of Eq.~\eqref{eq: JN theorem} were previously studied for ohmic devices \cite{Prober1992, Sukhorukov1999, Pozderac2021}, i.e. where $j_i(\mathbf{x}) = \sigma_{ij}(\mathbf{x}) E_j(\mathbf{x})$ holds locally. In non-equilibrium ohmic systems, it was shown that the Johnson-Nyquist theorem can be generalized to
\begin{align}
    \delta T_\text{JN}^\text{ohm} \equiv T_\text{JN}^\text{ohm} - T_0 = \frac{P R_\text{th}}{12}
    \label{eq: T_JN ohmic}
\end{align}
where $P$ is the input power and $R_\text{th}$ is the two-terminal thermal resistance (i.e. $R_\text{th} \propto 1/\kappa$, for $\kappa$ the thermal conductance). This result, perhaps surprisingly, is independent of the details of the particular two-terminal geometry and has been powerfully used in noise thermometry experiments \cite{Fong2012, Fong2013, Crossno2016, Talanov2021, Waissman2022, Talanov2024}.

However, with the increasing interest in thermal properties of non-ohmic systems, one cannot simply apply noise thermometry techniques using the ohmic Eq.~\ref{eq: T_JN ohmic}. This is dramatically demonstrated in Ref.~\onlinecite{Talanov2024}, where for a graphene sample in the Fermi liquid regime, they observed a decrease in the measured Johnson noise $\delta T_{\text{JN}}$ with an increase in magnetic field. A na\"ive application of the ohmic Eq.~\eqref{eq: T_JN ohmic} implies that the thermal resistance decreases with increasing magnetic field (i.e. negative thermal magnetoresistance). This in direct contradiction with expected theory, since a magnetic field generally increases path lengths and therefore increases resistances (see e.g. \onlinecite{Lucas2018}). We resolve this conundrum below, 
arguing that Eq.~\eqref{eq: T_JN ohmic} fails to account for hydrodynamic non-localities; we show that the decrease in $\delta T_{\text{JN}}$ is a reflection of the magnetic-field dependence of viscous heat dissipation in a Corbino geometry.

In this paper, we focus on hydrodynamic electron systems. Hydrodynamic behavior has been experimentally demonstrated in a number of materials, where the presence of viscosity invalidates a local Ohm's law \cite{Lucas2018, Narozhny2022, deJong1995, Muller2009, Torre2015, Levitov2016, Bandurin2016, Crossno2016, Guo2017, Kumar2017, Bandurin2018, Braem2018, Hui2020, Bergdyugin2019, Gallagher2019, Sulpizio2019, Jenkins2022, Ku2020, Vool2021, Aharon-Steinberg2022, Moll2016, Bachmann2022, Gooth2018, Gusev2018, Levin2018, Gusev2020, Shavit2019, Stern2022, Kumar2022, valentinis2023, Gall2023, Gall2023a, Hui2021, Hui2023b, Samaddar2021, Huang2023, Muller2008, Foster2009, Principi2015, Lucas2016, Zarenia2019, Zarenia2020, Robinson2021, Ahn2022, Li2022, Hui2023, Shavit2019, Stern2022, Kumar2022}. Graphene specifically has served as the prime candidate for hydrodynamic behavior due to its weak electron-phonon coupling, low disorder, and strong electron-electron interactions \cite{Lucas2018, Narozhny2022, deJong1995, Muller2009, Torre2015, Levitov2016, Bandurin2016, Crossno2016, Guo2017, Kumar2017, Bandurin2018, Braem2018, Hui2020, Bergdyugin2019, Gallagher2019, Sulpizio2019, Jenkins2022, Ku2020, Vool2021, Aharon-Steinberg2022, Moll2016, Bachmann2022, Gooth2018, Gusev2018, Levin2018, Gusev2020, Shavit2019, Stern2022, Kumar2022, valentinis2023, Gall2023, Gall2023a, Hui2021, Hui2023b}. 
Hydrodynamic Johnson noise was recently studied for the specific case of a two-terminal rectangular geometry with current-bias heating \cite{Hui2023}, giving
\begin{align}
    \delta T_\text{JN} = \frac{P R_\text{th}}{12}f
    \label{eq: JN general}
\end{align}
where $P$ is the input power, $R_\text{th}$ is the two-terminal thermal resistance, and $f$ is a non-universal function of geometry and of the Gurzhi length $\lambda = \sqrt{\nu/\gamma}$ for $\nu$ the kinematic viscosity and $\gamma$ the momentum-relaxation rate. In the ohmic limit, one recovers the $f=1$ result. This demonstrated that the ohmic Eq.~\eqref{eq: T_JN ohmic} \cite{Sukhorukov1999, Pozderac2021} generically requires a correction by $f$ in the presence of viscosity. Somewhat miraculously, for a rectangular geometry this function $f$ never causes a correction larger than $40\%$; in particular, this result placed the seminal measurements of Wiedemann-Franz violation \cite{Crossno2016} in graphene, which na\"ively used the $f=1$ ohmic result, on firm theoretical foundation.

While the rectangular geometry is sufficient to demonstrate violation of the ohmic $f=1$ result as shown in Eq.~\eqref{eq: JN general}, it critically misses subtleties about hydrodynamic flow in general geometries. This is strikingly displayed by the ``Corbino paradox'' \cite{Hui2022, Shavit2019, Stern2022, Kumar2022}, where a finite current locally dissipates heat despite the absence of a local electric field. In a circular-symmetric geometry, the radial current density is fixed to be $j_r \propto 1/r$ by current conservation. Since the Laplacian of the current density $\nabla^2 j_r \equiv 0$, there is zero viscous force throughout the bulk of the sample. By solving the (purely viscous) Stokes equation, this implies that electric field $\nabla \phi \equiv 0$ is identically zero in the bulk for a purely viscous sample. Despite the fact that the local electric field does no work, there is a non-zero viscous stress throughout the sample which leads to heat dissipation. Ultimately, heat dissipation implies that there must be a voltage drop between the two leads; thus, the Corbino paradox is resolved by the presence of sharp contact voltage drops at the boundary concomitant with zero electric field in the bulk, leading to an effective two-terminal resistance between the two metallic leads. In fact, this effect has recently been experimentally observed \cite{Kumar2022}. Crucially, the unusual vanishing of electric field in the bulk is a fundamental feature of viscous flow in a Corbino device.

For the purpose of noise measurements, the Corbino paradox poses yet another corollary paradox. The Shockley-Ramo theorem \cite{Shockley1938, Ramo1939, Yoder1996, Cavalleri1971, He2001, Song2014} states that for a velocity source $\mathbf{v}_s$, the outgoing current $I_\alpha^\text{SR}$ through contact $\alpha$ is given by
\begin{align}
    I_\alpha^\text{SR} = nq \int d^2 \mathbf{r} \nabla \phi_\alpha (\mathbf{r}) \cdot \mathbf{v}_s (\mathbf{r})
    \label{eq: SR theorem}
\end{align}
where $\phi_\alpha$ is the resulting potential distribution when the potential at contact $\alpha$ is set to unity and all other contacts are grounded. In what follows, we will suppress the position argument $\mathbf{r}$. Since a hydrodynamic Corbino device always has $\nabla \phi_\alpha = 0$, a na\"ive application of Eq.~\eqref{eq: SR theorem} gives $I_\alpha^\text{SR} = 0$ identically. This yields what we call the ``Corbino Shockley-Ramo Paradox'': a na\"ive and, as we show below, incorrect application of the Shockley-Ramo theorem implies that bulk velocity fluctuations would never result in measured current fluctuations at the contacts. This would unphysically imply that current noise fails to measure anything about the bulk Corbino device. Therefore, this Corbino Shockley-Ramo paradox indicates a fundamental problem of applying the Shockley-Ramo theorem to hydrodynamic electron systems, e.g. to the measurements of Ref.~\onlinecite{Talanov2024}.

In this work, we resolve the mysteries of negative thermal magnetoresistance of Ref.~\onlinecite{Talanov2024} and the Corbino Shockley-Ramo Paradox. To do this, we properly formulate the problem for current noise in a multi-terminal device of arbitrary geometry, generalizing the boundary condition of Ref.~\onlinecite{Shavit2019}. The Corbino Shockley-Ramo paradox is resolved by introducing the ``hydrodynamic Shockley-Ramo'' problem, analogous to original Shockley-Ramo problem for electrons in free space. We begin by showing that our formulation completely reproduces previous results in the ohmic limit. Then, we explicitly solve for the Johnson noise in a Corbino geometry under magnetic field. For current-bias heating, we find that the correction $f$ decreases with $B$ in the viscous limit (see Fig.~\ref{fig: f(B) normalized}). This provides a mechanism for the unusual experimental observation of Johnson noise suppression by magnetic field in Corbino devices \cite{Talanov2024}, which we argue is a novel qualitative signature of viscous hydrodynamics. Finally, we revisit the rectangular geometry and validate the previously unjustified ``Shockley-Ramo assumption'' made in Ref.~\onlinecite{Hui2023}.

\section{Mathematical Setup}

\subsection{Generic ohmic-Stokes equations}

\begin{figure}
    \centering
    \begin{subfloat}[][]{
        \includegraphics[width=.45\columnwidth]{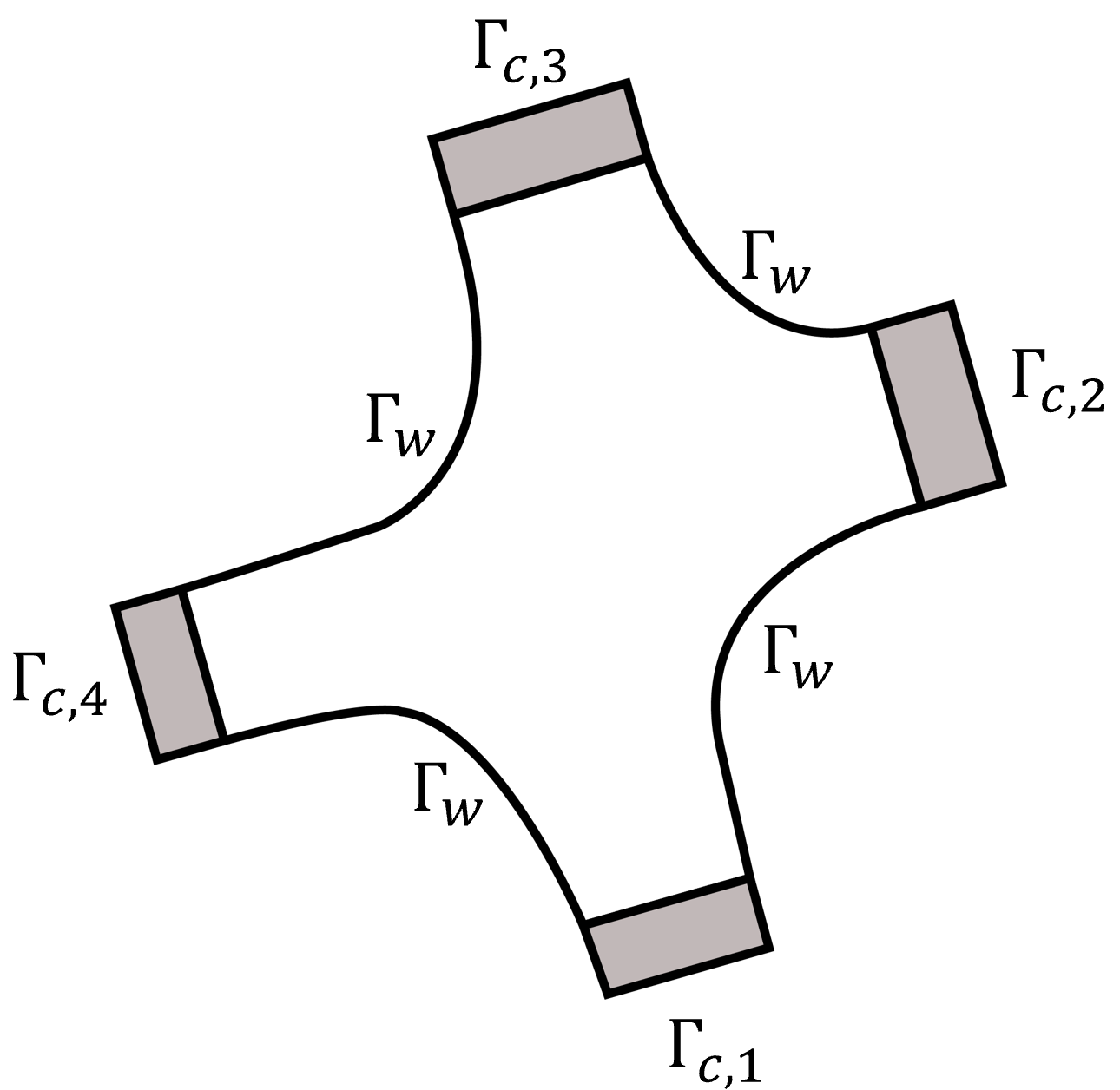}
        \label{fig: multiterminal geometry}
    }
    \end{subfloat}
    \quad
    \begin{subfloat}[][]{
        \includegraphics[width=.35\columnwidth]{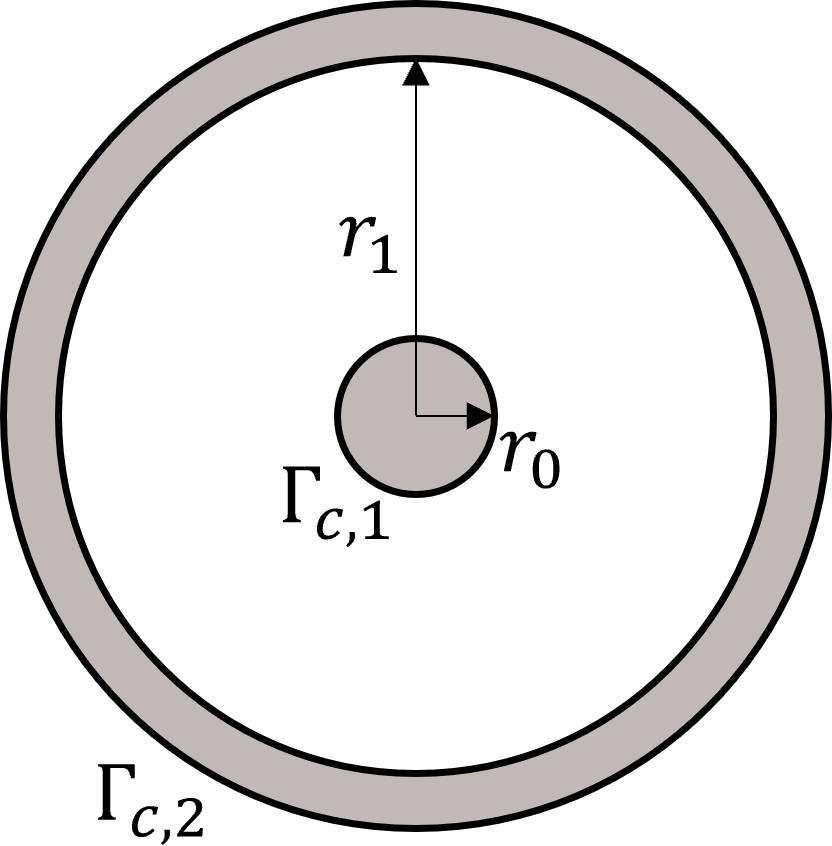}
        \label{fig: Corbino geometry}
    }    
    \end{subfloat}
    \caption{A plot of representative geometries. a) An arbitrary 4-terminal geometry. b) The Corbino geometry.}
    \label{fig: geometries}
\end{figure}

We begin by treating the ohmic-Stokes problem in full generality with a detailed discussion of the subtleties around the Corbino paradox \cite{Hui2022, Shavit2019, Stern2022}. This mathematical detour will provide a time-saving framework for the results to follow. The generic incompressible ohmic-Stokes equations on a connected domain $\Omega \subset \mathbb{R}^2$ (see Fig.~\ref{fig: geometries}) are given by
\begin{align}
    \mathcal{L}[\mathbf{v}] \equiv \gamma \mathbf{v} - \nu \nabla^2 \mathbf{v} - \frac{q}{m} \mathbf{v} \times \mathbf{B}  &= -\frac{q}{m}\nabla \phi +\mathbf{g}
    \label{eq: NS compact}
    \\
    \nabla\cdot \mathbf{v} &= 0
    \label{eq: cont}
\end{align}
where $\nu$ is the kinematic viscosity, $q$ the hydrodynamic charge, $m$ the hydrodynamic mass, $\mathbf{B} = B \mathbf{\hat{z}}$ the magnetic field, and $\mathbf{g}$ a given forcing term ($\nabla^2 \mathbf{v}$ is the \emph{vector} Laplacian). The linear operator $\mathcal{L}[\mathbf{v}]$ is used for notational convenience. In particular, we neglect convection; this is valid when the Reynolds numbers $\operatorname{Re} \equiv vL/\nu$ and $\operatorname{Re}_\gamma \equiv v/L\gamma$ are small, where $L$ is a characteristic length \cite{Hui2021}. To finish the specification of the PDE, one must supply boundary conditions. We assume that the boundary $\partial \Omega$ can be split into two distinct categories: $\Gamma_w$ corresponds to sample walls and $\Gamma_c = \cup_\alpha \Gamma_{c,\alpha}$ corresponds to the set of metallic contacts \footnote{Technically, we assume $\Gamma_w, \Gamma_c$ are open sets such that $\Gamma_w \cap \Gamma_c = \emptyset$ and $\partial\Omega = \overline{\Gamma_w} \cup \overline{\Gamma_c}$, where the overbar denotes closure. We take $\Gamma_{c,\alpha}$ to be the connected components of $\Gamma_c$. Finally, we assume the boundary is Lipschitz continuous; that is to say, piecewise segments are sufficiently smooth.}. We take the following BCs (boundary conditions)
\begin{align}
    \mathbf{v}|_{\Gamma_w} &= 0
    \label{eq: wall no-slip BC}
    \\
    \quad v_\parallel|_{\Gamma_c} &= 0
    \label{eq: contact no-slip BC}
    \\
    (\phi^c - \phi)|_{\Gamma_c} &= -\left.\frac{\sigma'_{nn}[\mathbf{v}]}{nq} \right|_{\Gamma_c}
    \label{eq: contact BC}
\end{align}
where $v_\parallel$ is the velocity tangent to the boundary, $\phi^c|_{\Gamma_{c,\alpha}} \equiv \phi^c_\alpha$ is a fixed external voltage on each of the contacts $\Gamma_{c,\alpha}$, $\sigma'_{ij}[\mathbf{v}] \equiv mn\nu (\partial_i v_j + \partial_j v_i)$ is the viscous (deviatoric) stress tensor with particle density $n$, and $\sigma'_{nn}[\mathbf{v}] \equiv \sigma'_{ij}[\mathbf{v}] n_i n_j$ for $n_i$ the outward normal unit vector. The normal component of the stress tensor $\sigma'_{nn}$ can be explicitly written as
\begin{align}
    -\left.\frac{\sigma'_{nn}[\mathbf{v}]}{nq} \right|_{\Gamma_c} = -\frac{2m\nu}{q} K(\hat{\mathbf{n}}) (\mathbf{v}\cdot \hat{\mathbf{n}}) \Big|_{\Gamma_c}
    \label{eq: stress tensor polar coordinates}
\end{align}
where $K(\hat{\mathbf{n}})$ is the signed extrinsic curvature relative to $\hat{\mathbf{n}}$ \footnote{In local polar coordinates, $K(\hat{\mathbf{n}}(\mathbf{x})) = \hat{\mathbf{n}}\cdot \hat{\mathbf{r}}/r$, where $r$ is the radius of the osculating circle at the point $\mathbf{x}$ on the boundary.}; this statement can be derived using incompressibility [Eq.~\eqref{eq: cont}] \cite{Barth2007, Shavit2019}. The no-slip conditions Eq.~(\ref{eq: wall no-slip BC}-\ref{eq: contact no-slip BC}) seem to be a reasonable approximation of experiment \cite{Sulpizio2019, Ku2020, Vool2021, Kumar2022, Jenkins2022} and are taken for simplicity. The BC of Eq.~\eqref{eq: contact BC} was first introduced in the electron hydrodynamics literature by Ref.~\onlinecite{Shavit2019}, though in the fluid dynamics literature it has a lengthy history \cite{Heywood1996, Barth2007, Bertoluzza2017} as researchers explored how to properly implement pressure-fixed BCs on arbitrary domains. In what follows, we will refer to the Ohmic-Stokes problem of Eqs.~(\ref{eq: NS compact}-\ref{eq: contact BC}) with the shorthand $\mathbb{S}(\mathbf{v}; \mathbf{g}, \phi^c_\alpha)$; this will allow us to alter the inputs for the bulk forcing $\mathbf{g}$ and the BCs $\phi^c_\alpha$ for this problem in a concise manner. 

The justification of Eq.~\eqref{eq: contact BC} requires physical arguments on the nature of the boundary $\Gamma_c$. At the interface, we expect current-density continuity. Furthermore, we assume the metallic contact only dissipates $-\nabla \phi^c \cdot \mathbf{j}$ power. Then, upon multiplication by the normal current density $j_n$, we see that Eq.~\eqref{eq: contact BC} is nothing but a continuity statement of energy-flux density across the boundary; Eq.~\eqref{eq: contact BC} says that energy transfer across the interface via the hydrodynamic flow is dissipationless \footnote{Concretely, the outgoing energy-flux from the bulk is given by $\sigma_{nn} v_n \equiv (nq\phi - \sigma_{nn}')v_n$, where $\sigma_{ij}$ is the total stress (since $v_\parallel = 0$). Eq.~\eqref{eq: contact BC} equates this to the incoming energy-flux to the contact $\phi^c j_n$.}. In this light, we can reframe the origin of the viscous ``contact voltage drop''. This voltage drop accounts for the ``viscous energy'' associated with maintaining a finite $\sigma'_{nn}$, which is only non-zero on locally-curved portions of the contact \cite{Shavit2019, Barth2007}. For flat contacts or in the ohmic limit, we recover the usual voltage-continuity BC $\phi^c|_{\Gamma_c} = \phi|_{\Gamma_c}$. This perspective of energy-flux continuity allows us to readily generalize the BC of Ref.~\onlinecite{Shavit2019} to multi-terminal ohmic-Stokes devices [Eq.~\eqref{eq: contact BC}] since ohmic dissipation and a magnetic field do not contribute to energy-flux.

It is important to ask whether the ohmic-Stokes problem $\mathbb{S}(\mathbf{v}; \mathbf{g}, \phi^c_\alpha)$ is well-posed, i.e. whether it has an existence-uniqueness (and regularity) result. On physical grounds, an ill-posed PDE implies that it is an improper description of reality. As an example of an ill-posed PDE, set $\gamma = B = 0$ and consider a Corbino geometry with the na\"ive BC of $\phi|_{\Gamma_c} = \phi^c|_{\Gamma_c}$ [instead of Eq.~\eqref{eq: contact BC}] and a constant voltage drop $V$ across the contacts. Then, the solutions are given by $v_r = \frac{I}{nq} \frac{1}{2\pi r}$ for arbitrary constant $I$ and $V\equiv 0$. This is the ``Corbino paradox'' \cite{Hui2022}: an arbitrary current is allowed to flow with no applied voltage drop despite the fact that each choice of $I$ dissipates a different amount of energy. Thus, this na\"ive BC is unphysical. More generally, it is known that this PDE with the na\"ive BC is ill-posed so long as $\Gamma_w = \emptyset$ \cite{Conca1994}. While the demonstration of existence-uniqueness with the BC of Eq.~\eqref{eq: contact BC} is beyond the scope of this paper, it is known that the vanilla Stokes equation ($B = \gamma = 0$) with these BCs does indeed enjoy existence-uniqueness \cite{Bertoluzza2017}. It is therefore natural to conjecture that the extension to the magnetic ohmic-Stokes case is also well-posed.

The primary quantity of interest in the ohmic-Stokes problem is the total outgoing current $I_\alpha$ through contact $\alpha$. If our problem is well-posed, then we can express the general solution $\mathbf{v} \equiv \mathcal{L}_v^{-1}[\mathbf{g}, \phi^c_\alpha]$ by a linear operator $\mathcal{L}_v^{-1}$ acting on the bulk forcing $\mathbf{g}$ and the set of contact voltages $\phi_\alpha^c$ \footnote{The inverse notation is motivated from the fact that the weak formulation of the ohmic-Stokes problem can be expressed as a linear operator equation for $\mathbf{v}$. Solving for $\mathbf{v}$ is then equivalent to finding the inverse operator, which we call $\mathcal{L}_v^{-1}$ \cite{Evans2010book, temam2016book}. In particular, $\mathcal{L}_v^{-1}$ is not quite the inverse of $\mathcal{L}$ due to the fact that $\phi$ may be non-zero to enforce incompressibility. We also note that second argument of $\mathcal{L}_v^{-1}$ takes the set of contact voltages $\{\phi_\alpha^c\}$, suppressing the set notation for readability.}. As a technical trick, we can choose (sufficiently smooth) weighting potentials $\psi_\alpha$ on $\Omega$ such that $\psi_\alpha|_{\Gamma_{c,\beta}} = \delta_{\alpha\beta}$ and $\sum_\alpha \psi_\alpha = 1$ \footnote{One can explicitly obtain such a weighting potential by, e.g., solving the Laplace equation for $\psi_\alpha$ with these BCs on $\Gamma_{c,\beta}$ and $\partial_n \psi_\alpha|_{\Gamma_w} = 0$. This corresponds to the potential profile on an ohmic device with contact $\Gamma_{c,\alpha}$ held at unit voltage and all others grounded.}. This partition of unity affords us nice simplifications. First, we note 
\begin{align}
    \mathcal{L}_v^{-1}\left[0,\overline{\phi^c_\alpha}\right] = -\frac{q}{m} \sum_\alpha  \overline{\phi^c_{\alpha}} \mathcal{L}_v^{-1}[\nabla \psi_\alpha,0]
\end{align}
via a redefinition of $\phi$, where $\overline{\phi^c_{\alpha}}$ are constants. Thus, the partition of unity allows us to express the applied voltage BC as a forcing term instead. Second, by multiplying Eq.~\eqref{eq: cont} by $\psi_\alpha$ and integrating by parts, we can express the outgoing current $I_\alpha$ and conductance $G_{\alpha\beta}$ as
\begin{align}
    I_{\alpha} =& G_{\alpha\beta} \overline{\phi^c_\beta} + nq \int dV \nabla \psi_\alpha \cdot \mathcal{L}_v^{-1}[\mathbf{g},\delta \phi^c_\beta]
    \label{eq: abstract current}
    \\
    G_{\alpha\beta} =& -\frac{nq^2}{m} \int dV \nabla \psi_\alpha \cdot \mathcal{L}_v^{-1}[\nabla \psi_\beta,0]
    \label{eq: abstract conductance}
\end{align}
where $\delta \phi^c_\alpha \equiv \phi^c_\alpha - \overline{\phi^c_\alpha}$ is the zero-mean component. Since we are primarily interested in $I_\alpha$, for our purposes Eq.~\eqref{eq: abstract current} constitutes the (abstract) solution to the ohmic-Stokes problem. While utilizing these expressions in full generality is manifestly just as difficult as solving the generic ohmic-Stokes problem for $\mathcal{L}^{-1}_v$, we can still obtain analytic results for some simple cases such as the Corbino.

\subsection{Hydrodynamic Shockley-Ramo Problem}

With preliminaries out of the way, we set up the hydrodynamic Shockley-Ramo problem: in a ohmic-Stokes sample, compute the total current through each contact induced by current sources embedded inside the sample bulk. The analogous problem was treated for the free-space \cite{Shockley1938, Ramo1939} and ohmic \cite{Yoder1996, Cavalleri1971, He2001, Song2014} cases, where there is a simple result known as the (generalized) Shockley-Ramo theorem. The problem is formulated as follows. We consider a given velocity source $\mathbf{v}_s$ which drives a total velocity $\mathbf{v}$ controlled by the following ohmic-Stokes PDE
\begin{align}
    \mathcal{L}[\mathbf{v} - \mathbf{v}_s] + \frac{q}{m} \nabla \phi =& 0
    \\
    \nabla \cdot \mathbf{v} =& 0
    \\
    \mathbf{v}|_{\Gamma_w} =& 0
    \\
    v_{\parallel}|_{\Gamma_c} =& 0
    \\
    - \phi|_{\Gamma_c} =&  -\frac{2m\nu}{q} K(\mathbf{\hat{n}}) (\mathbf{v}-\mathbf{v}_s)\cdot \hat{\mathbf{n}} \big|_{\Gamma_c}
    \label{eq: hydro SR contact BC}
\end{align}
where we have set $\phi^c = 0$. Physically, we interpret $\mathbf{v} - \mathbf{v}_s$ and $\phi$ as corresponding to the velocity and electrochemical potential of ``internal carriers'' of the sample; these carriers obey the ohmic-Stokes equations of motion and conspire to satisfy BC and incompressibility constraints. By linearity, we can rewrite this into the ohmic-Stokes problem $\mathbb{S}(\mathbf{v}; \mathcal{L}[\mathbf{v}_s], \frac{2m\nu}{q} K(\hat{\mathbf{n}}) v_{s,n})$. Using Eq.~\eqref{eq: abstract current}, we can write the solution as
\begin{align}
    I_\alpha^\text{SR} = nq \int dV \nabla \psi_\alpha \cdot \mathcal{L}_v^{-1}\left[\mathcal{L}[\mathbf{v}_s],\frac{2m\nu}{q} K(\hat{\mathbf{n}}) v_{s,n}\right]
    \label{eq: SR Solution}
\end{align}
This provides the proper generalization of the (ohmic) Shockley-Ramo theorem [Eq.~\eqref{eq: SR theorem}] for ohmic-Stokes devices. We see there are two effects contributing to the measured current $I_\alpha^\text{SR}$: a bulk forcing from the velocity source $\mathcal{L}[\mathbf{v}_s]$ and a boundary potential $(2m\nu/q) K(\hat{\mathbf{n}}) v_{s,n}$. Furthermore, we avoid the previous paradoxical Corbino conclusion of vanishing current for arbitrary velocity source due to the presence of $\psi_\alpha$; the electric potential $\phi_\alpha$, especially for the Corbino when $\phi_\alpha \equiv 0$, does not generally serve as a valid partition of unity $\psi_\alpha$.

Furthermore, to obtain correct results (e.g. recovering the Johnson-Nyquist theorem) we emphasize that the voltage BC [Eq.~\eqref{eq: hydro SR contact BC}] must be chosen correctly. The justification of Eq.~\eqref{eq: hydro SR contact BC} is quite subtle, as it amounts to the question of how to treat the velocity source $\mathbf{v}_s$ ``on the boundary.'' We argue as follows. Physically, we imagine the velocity source is close to but not quite on the boundary itself. Between the source and the boundary, there are no more sources so that $\mathbf{v}_s$ is continuous and divergence-free in this ``Knudsen'' region. Thus, the energy-flux transmitted by $\mathbf{v}_s$ is precisely given by the incompressible form of Eq.~\eqref{eq: contact BC}. We then take the limit of the source-boundary distance to zero to ``define'' the meaning of $\mathbf{v}_s |_{\Gamma_c}$. This $\mathbf{v}_s$ energy-flux contribution is subtracted in Eq.~\eqref{eq: hydro SR contact BC} to obtain the corresponding energy-flux statement for the internal electrochemical potential $\phi$.

\subsection{Noise problem formulation}

To compute the current-current correlator for Johnson noise, we proceed in two steps. First, we solve the hydrodynamic Shockley-Ramo problem as formulated above. This gives a relation between the (thermally-driven) bulk velocity fluctuations and current fluctuations as measured by the contacts at $t=0$. The result is used in the initial condition for the time-evolution equation of the correlation functions, which we dub the noise equation. Solving the noise problem in the $\omega \rightarrow 0$ limit, we can finally obtain the current-current correlator as a function of the local temperature $T(\mathbf{r})$. In particular, both of these problems reduce to an ohmic-Stokes problem; once formulated appropriately, one can apply the general expressions for current and conductance [Eq.~\eqref{eq: abstract current} and Eq.~\eqref{eq: abstract conductance}] to give the desired result.

To formulate the noise equation, we find it easier to work with the ``fluctuating states'' formulation rather than with Langevin forces \cite{landauv9, Hui2023}. We assume that the correlation functions obey the same time-dependent equations of motion as their non-equilibrium counterpart. Taking the Laplace transform of the time-dependent ohmic-Stokes PDE and Wick-rotating to obtain the two-sided Fourier transform, in the $\omega\rightarrow 0$ limit we have
\begin{align}
    \mathcal{L}[\langle \delta \mathbf{v} \delta I_{0,\beta} \rangle] =& -\frac{q}{m}\nabla \langle \delta \phi \delta I_{0,\beta} \rangle + 2\langle \delta \mathbf{v}_s \delta I_{0,\beta} \rangle
    \\
    \nabla \cdot \langle \delta \mathbf{v} \delta I_{0,\beta} \rangle =& 0
\end{align}
where we have suppressed the frequency argument $\delta \mathbf{v} \equiv \delta \mathbf{v}(\omega\rightarrow 0)$ of the velocity fluctuations and have ignored thermal density fluctuations to obtain incompressibility \footnote{Thermal density fluctuations are suppressed by the thermal Mach number $\sqrt{k_B T/(mc^2)}$, where $c$ is the speed of sound \cite{landauv9}. A rough estimate with $m = m_e$ the electron mass at $T=300K$ and $c \sim v_F \sim 10^6 m/s$ gives $\text{Ma}_\text{th} \sim .07$, so we expect thermal density fluctuations to be a subleading effect.}. The noise source $\langle \delta \mathbf{v}_s \delta I_{0,\beta} \rangle$ arises from the initial conditions \footnote{A factor of two arises from the fact $\langle \delta \mathbf{v}(\omega) \delta I_{0,\beta}\rangle = 2\langle \delta v(s) \delta I_{0,\beta} \rangle$ upon conversion to the two-sided Fourier transform because the correlation function $\langle \mathbf{v}(t) I_{0,i} \rangle$ is pure real and time-reversal even.}, where $\delta \mathbf{v}_s$ is the source velocity and $\delta I_{0,\beta} = \delta I_\beta(t=0)$ is the initial current through contact $\Gamma_{c,\beta}$  \footnote{While it is mnemonically convenient to identify $\delta \mathbf{v}_s$ with $\delta \mathbf{v} (t=0)$, this is not actually correct; $\delta \mathbf{v}_s$ need not obey incompressibility.}. For BCs, we set $\delta \phi^c = 0$ take the same no-slip and energy-flux continuity BCs as Eq.~(\ref{eq: wall no-slip BC}-\ref{eq: contact BC}) on the fluctuations $\delta\mathbf{v}$ and $\delta \phi$. Thus, the noise equation reduces to solving the ohmic-Stokes problem $\mathbb{S}(\langle\delta \mathbf{v} \delta I_{0,\beta} \rangle; 2\langle \delta \mathbf{v}_s \delta I_{0,\beta} \rangle, 0)$ for the current-current correlator $\langle \delta I_{0,\alpha} \delta I_{0,\beta} \rangle$. Explicitly, the solution reads
\begin{align}
    \langle \delta I_{0,\alpha} \delta I_{0,\beta} \rangle =& nq \int dV \nabla \psi_\alpha \cdot \mathcal{L}_v^{-1}[2\langle \delta \mathbf{v}_s \delta I_{0,\beta} \rangle,0]
\end{align}
To evaluate the initial condition $\langle \delta \mathbf{v}_s \delta I_{0,\beta} \rangle$, we utilize the result of the hydrodynamic Shockley-Ramo problem [Eq.~\eqref{eq: SR Solution}] to compute $\delta I_{0,\beta} = I^\text{SR}_\beta[\delta\mathbf{v}_s]$ and then apply the thermodynamic relation 
\begin{align}
\langle \delta v_{s,i}(\mathbf{r}) \delta v_{s,j} (\mathbf{r}') \rangle = \frac{k_B T(\mathbf{r})}{mn} \delta(\mathbf{r} - \mathbf{r}') \delta_{ij} 
\label{eq: equipartition theorem}
\end{align}
with temperature profile $T(\mathbf{r})$. We remark that this formulation doesn't require a thermal noise source; the noise input is defined by the choice of $\langle \delta v_{s,i}(\mathbf{r}) \delta v_{s,j} (\mathbf{r}') \rangle$.

To make concrete experimental predictions, we will primarily be interested in non-equilibrium temperature profiles generated by a current-bias heating \cite{Fong2012, Crossno2015, Crossno2016, Talanov2021, Waissman2022}. Solving for $T(\mathbf{r})$ proceeds in two steps. First, we solve the ohmic-Stokes PDE $\mathbb{S}(\mathbf{v}; 0, \phi^c)$ with applied voltages $\phi^c$ to obtain the velocity profile. This is used to compute the local heating $p[\mathbf{v}]$ for use in the heat equation
\begin{align}
    \kappa \nabla^2 T =& -p[\mathbf{v}] \equiv - \rho \gamma v^2 - \frac{1}{2\eta} (\sigma'_{ij}[\mathbf{v}])^2
    \label{eq: heat equation full}
    \\
    \partial_n T|_{\Gamma_w} =& 0
    \\
    T|_{\Gamma_c} =& T_0
\end{align}
where $\nabla^2 T$ is the \emph{scalar} Laplacian, $\kappa$ is the thermal conductivity, $\rho = mn$ is the mass density, and $\eta = mn\nu$ is the dynamic viscosity. We have neglected thermal advection for simplicity, which is valid when the thermal diffusion across the sample $\kappa/(\rho c_p L) \gg v$ is much faster than the hydrodynamic flow. For BCs, we assume that the sample walls are insulating and fix all contacts to be held at the ambient temperature $T_0$. We remark that the heating $q[\mathbf{v}]$ consists of two components - the usual ohmic Joule heating term as well as a viscous heating term; in particular, $q \neq \mathbf{E} \cdot \mathbf{J}$ as vividly demonstrated by the Corbino paradox \cite{Shavit2019, Hui2022}.

\section{Examples}

\subsection{Ohmic limit}

We begin by warming up on the ohmic limit $\nu \rightarrow 0$ \footnote{In this limit, the no-slip BCs on $\mathbf{v}_\parallel$ become spurious due to the loss of the derivative term; boundary layers are allowed to be arbitrarily thin so ohmic flow obeys no-slip in a discontinuous way.}. The ohmic-Stokes equation [Eq.~\eqref{eq: NS compact}] simplifies immensely because the differential operator $\mathcal{L} \propto \rho_{ij}$ reduces to the resistivity $\rho_{ij}$ and is easily inverted. Furthermore, there is a natural choice for $\psi_\alpha \equiv \phi_\alpha$, where $\phi_\alpha$ is the potential when setting $\phi^c_\beta|_{\Gamma_{c,\alpha}} = \delta_{\alpha\beta}$ to be unity only on $\Gamma_{c,\alpha}$. First, from the general expression for conductance [Eq.~\eqref{eq: abstract conductance}] we recover \cite{Sukhorukov1999}
\begin{align}
    G_{\alpha\beta} =& - \int dV \sigma_{ij} (\partial_i \phi_\alpha) (\partial_j \phi_\beta)
\end{align}
where we used the fact that $\int dV \nabla \phi \cdot \nabla \phi_\alpha = 0$ if $\phi|_{\Gamma_{c,\alpha}} = 0$. Next, we treat the ohmic Shockley-Ramo problem $\mathbb{S}(\mathbf{v}; \mathcal{L}[\mathbf{v}_s], 0)$. Using the same fact, the general expression for current [Eq.~\eqref{eq: abstract current}] gives for the outgoing current
\begin{align}
    I_\alpha^\text{SR} = nq\int dV \nabla \phi_\alpha \cdot \mathbf{v}_s
    \label{eq: ohmic I^SR}
\end{align}
Thus, we have recovered the ohmic Shockley-Ramo theorem [Eq.~\eqref{eq: SR theorem}] \cite{Shockley1938, Ramo1939, Yoder1996, Cavalleri1971, He2001, Song2014}. Finally, we solve the noise problem $\mathbb{S}(\langle\delta \mathbf{v} \delta I_{0,\beta} \rangle; 2\langle \delta \mathbf{v}_s \delta I_{0,\beta} \rangle, 0)$. From the Shockley-Ramo problem, we use Eq.~\eqref{eq: ohmic I^SR} for $\delta I_{0,\beta} = I^\text{SR}_\beta [\delta \mathbf{v}_s]$. Then, using the equipartition theorem Eq.~\eqref{eq: equipartition theorem} we find that $2\langle \delta v_{s,i} \delta I_{0,\beta} \rangle = - 2k_B T(\mathbf{r}) \frac{q}{m}\partial_i \phi_\beta$. Again repeating the same procedure, we find
\begin{align}
    \langle \delta I_\alpha \delta I_{0,\beta} \rangle = -2k_B \int dV T(\mathbf{r}) \sigma_{ij} (\partial_i \phi_\alpha) (\partial_j \phi_\beta)
\end{align}
recovering the result of Ref.~\cite{Sukhorukov1999} in the case of thermal noise. For $T = T_0$ we find that $\langle \delta I_\alpha \delta I_{0,\beta} \rangle = -2 k_B T_0 G_{\alpha\beta}$, recovering the Johnson-Nyquist theorem. From this point, it is manifest that we completely reproduce previous ohmic results from our formulation \cite{Sukhorukov1999, Pozderac2021}, including the two-terminal result \cite{Pozderac2021}
\begin{align}
    \delta T_{JN} = \frac{P R_\text{th}}{12}
    \label{eq: ohmic JN result}
\end{align}
As previously mentioned, this is Eq.~\eqref{eq: JN general} with $f=1$. 

\subsection{Corbino}

We now turn to a Corbino geometry with contacts on inner radius $r_0$ and outer radius $r_1$. In particular, the boundary only consists of contacts (see Fig.~\ref{fig: Corbino geometry}). As we will see, this geometry is particularly nice and admits an exact analytic solution. Instead of directly solving for $\mathcal{L}_v^{-1}$ and using the general expression for current [Eq.~\eqref{eq: abstract current}], we will obtain the total current by performing integration tricks to avoid solving for $\phi$. For computing the total current or heating density, we only need deal with angular-averaged quantities. In orthonormal polar $(r,\theta)$ coordinates, the ohmic-Stokes problem [Eqs.~(\ref{eq: NS compact}-\ref{eq: contact BC})] reads
\begin{align}
    \int dr \Big( L[\overline{v}_r] + \omega_c \overline{v}_\theta \Big) - 2\nu \left.\frac{\overline{v}_r}{r}\right|_{r_0}^{r_1}=& \frac{q}{m} V + \int dr \overline{g}_r
    \label{eq: NS radial}
    \\
    L[\overline{v}_\theta] =& \overline{g}_\theta + \omega_c \overline{v}_r
    \label{eq: NS angular}
    \\
    \overline{v}_r =& \frac{I}{nq} \frac{1}{2\pi r}
    \label{eq: v_r solution}
    \\
    \overline{\phi^c} - \overline{\phi} |_{\Gamma_c} =& -\frac{m\nu}{q} \left.\frac{\overline{v}_r}{r}\right|_{\Gamma_c}
\end{align}
with the no-slip BC $v_\theta|_{\Gamma_c} = 0$, where $\omega_c = qB/m$ is the cyclotron frequency, the linear operator $L[v] \equiv \gamma - \nu\left[\partial_r^2 v + \frac{1}{r}\partial_r v -\frac{v}{r^2}\right]$, and $\overline{X} = \int \frac{d\theta}{2\pi}X$ denotes an angular average. In what follows, we drop the overbar and implicitly assume all quantities are angular-averaged. Notice that incompressibility immediately gives us the solution for $v_r$, where $I$ is the constant radial current. With the no-slip BCs, Eq.~\eqref{eq: NS angular} decouples and can directly be solved for $v_\theta$. We denote the solution by $v_\theta = L^{-1}[g_\theta] + \omega_c L^{-1}[v_r]$ with the linear operator $L^{-1}$. Thus, Eq.~\eqref{eq: NS radial} gives
\begin{align}
    IR =& V + \frac{m}{q}\int dr \left(g_r - \omega_c L^{-1} [g_\theta]\right)
    \label{eq: abstract Corbino current}
    \\
    R =& \frac{m}{nq^2}\int dr (L + \omega_c^2 L^{-1})\left[\frac{1}{2\pi r}\right] - \frac{m\nu}{nq^2} \frac{1}{\pi} \left.\frac{1}{r^2}\right|_{r_0}^{r_1}
    \label{eq: abstract Corbino resistance}
\end{align}
where $V = -\phi^c|_{r_0}^{r_1}$ is the (angular-averaged) applied voltage \footnote{In particular, there is no $\delta\phi^c$ contribution to current (see Eq.~\eqref{eq: abstract current}) since the problem decouples under Fourier transform on $\theta$.}. Since the $L^{-1}$ operator is involved and not particularly enlightening, we will defer its evaluation. 

Turning to the hydrodynamic Shockley-Ramo problem $\mathbb{S}(\mathbf{v}; \mathcal{L}[\mathbf{v}_s], \frac{2m\nu}{q}K(\hat{\mathbf{n}}) v_{s,n})$, we use the expression for current [Eq.~\eqref{eq: abstract Corbino current}] to find that
\begin{align}
    I^\text{SR} R =& -\frac{2m\nu}{q} \left.\frac{v_{s,r}}{r}\right|_{r_0}^{r_1} + \frac{m}{q}\int dr (L + \omega_c^2 L^{-1})[v_{s,r}] 
    \nonumber
    \\
    &\phantom{-\frac{2m\nu}{q}} - \frac{m\omega_c}{q} \int dr (L^{-1} L - 1)[v_{s,\theta}]
    \label{eq: Corbino SR solution}
\end{align}
We see explicitly that the hydrodynamic result is quite different from the ohmic Shockley-Ramo theorem [Eq.~\eqref{eq: SR theorem}]. We remark that the operator $L^{-1} L - 1$ only depends on the boundary data $v_{s,\theta}|_\Gamma$; the corresponding integral will vanish for the noise problem since the noise input vanishes on the boundary.

Then, we solve the noise problem $\mathbb{S}(\langle\delta \mathbf{v} \delta I_{0,\beta} \rangle; 2\langle \delta \mathbf{v}_s \delta I_{0,\beta} \rangle, 0)$. After some algebra, we find the Johnson noise temperature to be
\begin{align}
    T_\text{JN} = T_0 +  \frac{m}{nq^2}\frac{1}{R} \int dr (L + \omega_c^2 L^{-1})\left[\frac{\delta T(r)}{2\pi r}\right]
    \label{eq: Corbino T_JN}
\end{align}
This yields a surprisingly simple result for the Johnson noise, arising from the fact that the equations in this problem decouple. We remark that the voltage BC [Eq.~\eqref{eq: hydro SR contact BC}] was critical in recovering the equilibrium Johnson-Nyquist theorem, as it provides the nontrivial boundary term in Eq.~\eqref{eq: Corbino SR solution}. At this stage, we see two explicit effects of a finite magnetic field which affect the Johnson noise. The first is to alter the magnetoresistance $R(B)$, and the second arises from altering the averaging procedure (i.e. the $\omega_c^2 L^{-1}$ term); while in the ohmic limit these effects cancel, this is not generally true in the hydrodynamic problem. Eq.~\eqref{eq: Corbino T_JN} is valid for arbitrary temperature distributions $T(r)$; to get a better physical sense of Eq.~\eqref{eq: Corbino T_JN}, below we specialize temperature profiles arising from current-bias heating.

\begin{figure}
    \centering
    \includegraphics[width=.95\columnwidth]{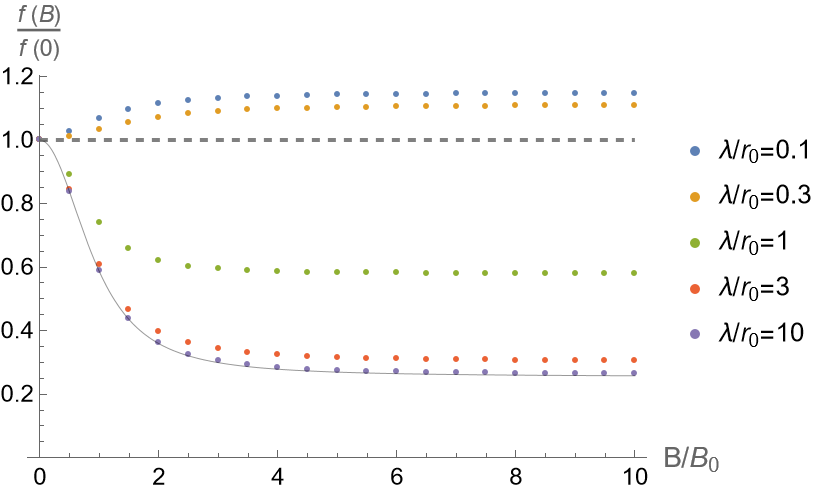}
    \caption{A plot of $f(B)/f(0)$ for $r_1/r_0 = 4.5$ in the Corbino geometry. The dashed line corresponds to the ohmic $\lambda = 0$ value, while the solid thin line corresponds to the viscous $\lambda \rightarrow \infty$ value. These curves are plotted against $B/B_0$, where the scale $B_0 = (m/q) (\gamma + \nu/r_0^2)$ is different for each curve.}
    \label{fig: f(B) normalized}
\end{figure}

\begin{figure}
    \centering
    \includegraphics[width=.8\columnwidth]{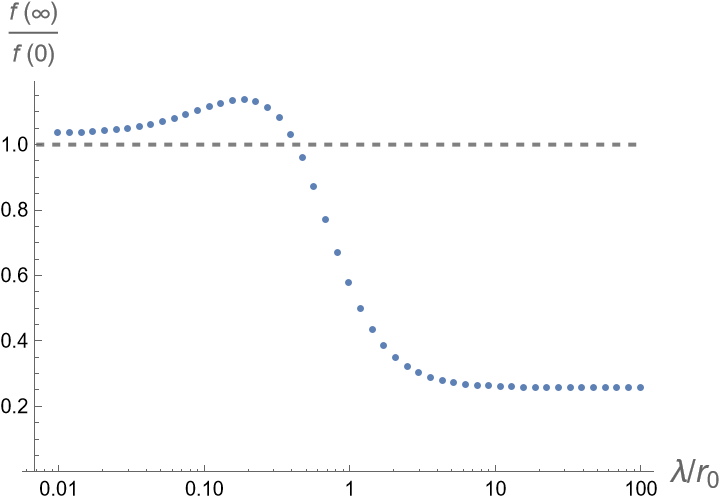}
    \caption{A plot of the limiting ratio $f(\infty)/f(0)$ as a function of $\lambda$ for $r_1/r_0 = 4.5$. The gray line denotes the ohmic $\lambda = 0$ value of $f(\infty)/f(0) = 1$. At low $\lambda$, we see that it at first increases, then dramatically drops around $\lambda/r_0 \sim 1$. This can be understood as a crossover to viscous flow and viscous heating in the limit of large $\lambda$, where the Johnson noise decreases since heating is less effective. We remark that there are numerical errors at low $\lambda$ due to the nearly discontinuous falloff of the $v_\theta$ profile near $r_0$ and $r_1$ (it is strictly discontinuous at the boundary in the ohmic limit).}
    \label{fig: f(inf)}
\end{figure}

To concretely evaluate the Johnson noise, we solve two PDEs: one for $L^{-1}$ and one for $\delta T(r)$ \footnote{Analytic results are available in the viscous $\lambda \rightarrow \infty$ limit, see the appendix.}. Defining $f$ using Eq.~\eqref{eq: JN general}, in Fig.~\ref{fig: f(B) normalized} we plot the numerical results \footnote{We numerically solve the PDEs by truncating the exact Fourier-Bessel type series solution. These solutions exist since $L^{-1}$ is given by the solution of a Sturm-Liouville ODE (in fact, an inhomogenous Bessel equation).} for $\frac{f(B)}{f(0)}$ against $B$ for a fixed aspect ratio $r_1/r_0 = 4.5$ and various Gurzhi lengths $\lambda \equiv \sqrt{\nu/\gamma}$. The Gurzhi length $\lambda$ can be physically interpreted as the length-scale over which momentum is approximately conserved, or equivalently below which viscous effects dominate. We see that $\frac{f(B)}{f(0)}$ weakly increases on the ohmic side $\lambda \ll r_0$, while it strongly decreases on the viscous side $\lambda \gg r_0$. Furthermore, we see that these curves saturate in the $B\rightarrow \infty$ limit due to the fact the flow profile reaches a limiting shape with magnetic field. In Fig.~\ref{fig: f(inf)}, we plot the asymptotic value $\frac{f(\infty)}{f(0)}$ as a function of $\lambda$ to more clearly depict the $\lambda$-dependence of the correction function $f$. 

We can qualitatively understand this behavior as follows, tracing through Fig.~\ref{fig: f(inf)}. For small $\lambda$, viscous corrections to flow are weak but a viscous heating channel emerges; this increases the total heating, and thus increases $T_\text{JN}/P \propto f(B)$. As one continues increasing $\lambda$, viscous corrections to flow begin to dominate and the ohmic heating is suppressed. Under a magnetic field, these flow corrections cause heating to be concentrated near the contacts; with a spiraling flow, the flow gradients and thus viscous heating are maximal at the boundaries because of the no-slip condition. This reduces the heating effectiveness and decreases $T_\text{JN}/P$. Thus, we predict a $B$-field suppression of $T_\text{JN}$ in the viscous hydrodynamic regime. This result matches the experimental observations in Ref.~\cite{Talanov2024}. 

We remark that this suppression of $T_\text{JN}$ with $B$-field is challenging to explain in an ohmic or ballistic regime. In the ohmic limit, where $f=1$, this would imply $R_\text{th}$ increases with $B$-field. However, in a conventional Boltzmann picture, a magnetic field curves the path of electrons. This increases path lengths, thus increasing scattering and increasing electrical and thermal resistance. In the ballistic limit, electrons do not thermalize nor significantly heat in the sample bulk; Johnson noise for ballistic samples primarily proceeds from the well-thermalized ohmic contacts. Under magnetic field, we still do not expect significant contribution to the Johnson noise from the bulk and therefore rule out $T_\text{JN}$ suppression in this regime \cite{Talanov2024}. We therefore conclude that $T_\text{JN}$ suppression by magnetic field is a strong signature of viscous hydrodynamic behavior.

\subsection{Rectangular geometry}

Finally, we revisit the noise problem for the rectangular geometry at $B=0$ as considered in Ref.~\onlinecite{Hui2023}. We take $\Omega = [0,\ell]\times [0,h]$ with contacts $\Gamma_c$ at $x=0,\ell$. We again compute the total current by clever integration tricks. Averaging over the $x$-component of the ohmic-Stokes equation [Eq.~\eqref{eq: NS compact}], we find
\begin{align}
    (\gamma - \nu \partial_y^2) \left[\frac{1}{\ell}\int dx v_x\right]  =& -\frac{q}{m\ell} V + \frac{1}{\ell}\int dx g_x
    \label{eq: rectangle eom}
\end{align}
where $V = \phi^c|_{x=0}^{x=\ell}$ and we have used $\partial_x v_x |_{\Gamma_c} = -\partial_y v_y |_{\Gamma_c} = 0$ from incompressibility and the no-slip BCs. Thus, we have reduced the problem to an ODE (ordinary differential equation) with vanishing BCs for $\int dx v_x$ on the $y$-boundaries. Fourier techniques allow us to solve for $\int dx v_x$, and by extension $I = (1/\ell) \int dV v_x$ with basis $\sin (k\pi y/h)$. Therefore, we have
\begin{align}
    I =& GV + \frac{nqh}{\gamma}\sum_{k=1}^\infty \frac{1}{1 + \left(\frac{k\pi \lambda}{h}\right)^2} \frac{1-\cos k \pi}{k \pi} \hat{g}_{x;0,k}
    \label{eq: rectangle current solution}
    \\
    G=& \frac{h\sigma_D}{\ell} \left(1 - \frac{2\lambda}{h} \tanh \frac{h}{2\lambda}\right)
\end{align}
where $\sigma_D = nq^2/(m\gamma)$ is the Drude conductivity, $\hat{g}_{x;0,k} \equiv \frac{2}{h\ell}\int dV g_x \sin \frac{k\pi y}{h}$ is the corresponding Fourier component, and $\lambda \equiv \sqrt{\nu/\gamma}$ is the Gurzhi length.

We now turn to the solution of the noise problem. As before, we first treat the hydrodynamic Shockley-Ramo problem $\mathbb{S}(\mathbf{v}; \mathcal{L}[\mathbf{v}_s], 0)$. The solution is given from Eq.~\eqref{eq: rectangle current solution} with $\mathbf{g} = \mathcal{L}[\mathbf{v}_s]$ and $V=0$, which reads
\begin{align}
    I^\text{SR} &= nq \int dV \frac{v_{s,x}}{\ell} 
    \nonumber
    \\
    & \phantom{\qquad} - nqh\sum_{k=1}^\infty \frac{\int  [\partial_x v_{s,x}]_0^\ell \sin \frac{k \pi y}{h} dy}{\frac{1}{\lambda^2} + \frac{k^2 \pi^2}{h^2}} \frac{1-\cos k \pi}{k \pi}
    \label{eq: SR rectangle}
\end{align}
where the second term arises from direct integration of $\nu \partial_x^2$. Thus, the second term appears as a correction to the ohmic Shockley-Ramo theorem in this geometry. \footnote{Note that $\mathbf{v} \neq \mathbf{v}_s$ (and $I^\text{SR} \neq I_s$) even though the equation of motion is $\mathcal{L}[\mathbf{v}] = \mathcal{L}[\mathbf{v}_s]$, as evidenced by the second term; this leads to a correction to the ohmic Shockley-Ramo result. This arises because $\mathbf{v}_s$ is a source current, and need not obey the constraints on $\mathbf{v}$ like boundary conditions or incompressibility; the electric potential $\phi$ redistributes and screens the forces due to $\mathbf{v}_s$ to satisfy the constraints on $\mathbf{v}$.}

To better understand the result Eq.~\eqref{eq: SR rectangle}, we explicitly write the ohmic-Stokes equation [Eq.~\eqref{eq: rectangle eom}] with our inputs.
\begin{align}
    (\gamma - \nu \partial_y^2)\int dx v_x = (\gamma - \nu \partial_y^2) \int dx v_{s,x} - \nu \left[\partial_x v_{s,x}\right]_{0}^{\ell}
\end{align}
By linearity, we can treat each term on the right-hand side separately. For the first term, by uniqueness of the ODE solution it can be quickly deduced that $\int dx v_x = \int dx v_{s,x}$. This directly gives the first term of Eq.~\eqref{eq: SR rectangle}, which corresponds to the ohmic Shockley-Ramo result. However, $x$-variations in $v_{s,x}$ also generate a viscous force, leading to the second term on the right-hand side. This does not have a simple counterpart for $v_x$ on the left-hand side due to incompressibility. Therefore, there is an additional contribution to current given by the second term of Eq.~\eqref{eq: SR rectangle} which augments the ohmic Shockley-Ramo result. In particular, this term is only non-zero if $\partial_x v_{s,x}$ is non-periodic along $x$.

We note that Eq.~\eqref{eq: SR rectangle} only recovers the ohmic Shockley-Ramo theorem [Eq.~\eqref{eq: SR theorem}] in the $\lambda \rightarrow 0$ limit due to the non-trivial second term ($\nabla \phi_\alpha = \pm 1/\ell$ for this geometry). However, the second term only depends on the boundary, so the noise input
\begin{align}
    \langle \delta v_{s,x} (\mathbf{r}) \delta I_0 \rangle = \frac{q}{m} \frac{1}{\ell} k_B T(\mathbf{r})
\end{align}
does indeed behave as if it obeys the ohmic Shockley-Ramo theorem for this geometry (recall we use $\delta I_0 = I^\text{SR}[\delta \mathbf{v}_s]$). This fills in a subtle unjustified assumption used in Ref.~\onlinecite{Hui2023}. For completeness, solving the noise problem gives a Johnson noise temperature
\begin{align}
    T_\text{JN} = \frac{R}{R_\text{ohm}} \sum_{k=1}^\infty \frac{1}{1 + \left(\frac{k\pi \lambda}{h}\right)^2} \frac{1-\cos m \pi}{m \pi} \hat{T}_{0,k}
\end{align}
where $R_\text{ohm} \equiv \ell/(h\sigma_D)$, recovering the result of Ref.~\onlinecite{Hui2023}. One can verify that for $T = T_0$ one indeed finds $T_\text{JN} = T_0$ as a sanity check.

\section{Conclusion}
In this paper, we have resolved the subtleties of the Corbino Shockley-Ramo paradox by correctly formulating the hydrodynamic Shockley-Ramo problem. This allows us to properly formulate the equations for hydrodynamic current noise in arbitrary multiterminal geometries and generalize previous ohmic results \cite{Prober1992, Sukhorukov1999, Pozderac2021}; our results enable proper interpretation of Johnson noise thermometry measurements in experimental setups \cite{Crossno2016, Waissman2022, Talanov2024}. We utilize this formulation for a Corbino geometry under magnetic field, where we find a strong signature of viscous heating in hydrodynamic flow: suppression of $T_\text{JN}$ by magnetic field (see also Ref.~\onlinecite{Talanov2024}). We also validate a previously unjustified Shockley-Ramo assumption in Ref.~\onlinecite{Hui2023} for a rectangular geometry.

While we have primarily focused on a Galilean-invariant fluid, our results can be directly extended to a Dirac fluid where both $n$-type and $p$-type carriers coexist (e.g., near the charge-neutral point in graphene), as discussed in Ref.~\onlinecite{Hui2023}. A Dirac fluid has an additional zero-momentum mode which can relax momentum through electron-hole scattering. This only changes the value of momentum-relaxation parameter $\gamma$ while the linear-response equations of motion remain untouched. Thus, our results continue to hold for this case.

We remark that hydrodynamic materials may prove ideal as highly sensitive noise thermometers due to the tunability of the Lorenz ratio and of the correction factor $f$ \cite{Hui2023}. Clever non-traditional geometries \cite{Waissman2022} have already been utilized to great effect to measure thermal properties with Johnson noise techniques. We leave further exploration of geometrical effects on noise in hydrodynamic devices and its optimization for noise thermometry applications to future work.

\textbf{Acknowledgements} -- I deeply thank Brian Skinner, Philip Kim, Jonah Waissman, and Artem Talanov for an exciting related collaboration and illuminating in-depth discussions. I also thank Gregory Falkovich and Justin Song for helpful discussions.

\bibliography{biblio}

\appendix
\onecolumngrid
\section{Corbino Calculational Details}

To evaluate $T_{JN}$ requires us to evaluate $L^{-1}$; i.e., we must solve $L[u] = g$ with homogeneous $u|_{\Gamma_c} = 0$ BCs \footnote{Similar to the rectangular case, this admits a generalized Fourier solution with the help of Sturm-Liouville theory; $L$ is diagonalized by the Bessel functions $J_1$ and $Y_1$.}. In the viscous $\gamma \rightarrow 0$ limit, we can integrate the ODE directly to obtain an analytic result,
\begin{align}
    \lim_{\gamma \rightarrow 0} L^{-1}[g] =& v_p(r) -\frac{r_1 v_p(r_1) - r_0 v_p(r_0)}{r_1^2 - r_0^2} r 
    \nonumber
    \\
    &+ \frac{r_0^2 r_1^2}{r_1^2 - r_0^2} \left(\frac{v_p(r_1)}{r_1} - \frac{v_p(r_0)}{r_0}\right) \frac{1}{r}
    \\
    v_p(r) \equiv & -\frac{1}{\nu} \frac{1}{r}\int^{r} dr' r' \int^{r'} dr'' g(r'')
\end{align}
where the particular solution $v_p$ is defined via antiderivatives; to fix the integration constants, one can arbitrarily fix the lower integration bound.

We also need to evaluate the temperature profile $T(r)$ under current-bias heating. Under a constant voltage bias $V$, the problem is rotationally symmetric; all quantities are equivalent to their angular average. With the help of $L^{-1}$, we know $\mathbf{v}$ from Eq.~\eqref{eq: NS angular} and Eq.~\eqref{eq: v_r solution}. This allows us to evaluate the heating density $q$ with Eq.~\eqref{eq: heat equation full}. Furthermore, rotational symmetry allows us to solve the heat equation by direct integration. The temperature profile is given by
\begin{align}
    \delta T(r) =& T_p(r) - T_p|_{r_0}^{r_1}\frac{\ln \frac{r}{r_0}}{\ln\frac{r_1}{r_0}} -T_p(r_0)
    \\
    T_p =& - \frac{1}{\kappa}\int^r dr' \frac{1}{r'} \int^{r'} dr'' r'' q(r'')
\end{align}
where again $T_p$ is a particular solution defined by antiderivatives.

We summarize the explicit results in the viscous limit $\gamma\rightarrow 0$ here:
\begin{align}
    v_\theta =& \frac{I}{nq} \frac{\omega_c}{2\pi\nu} \left[c_1 r + c_2\frac{1}{r} - \frac{1}{2} r\ln r \right]
    \\
    q=& \frac{m\nu}{nq^2} \frac{I^2}{(2\pi)^2}\left[\frac{4}{r^4} + \left(\frac{\omega_c}{2\nu}\right)^2 \left(1+\frac{4c_2}{r^2}\right)^2\right]
    \\ 
    T =& \frac{m\nu}{nq^2\kappa} I^2 \frac{1}{(2\pi)^2} \Bigg[ \left(\frac{1}{r_0^2}-\frac{1}{r^2} - \left(\frac{1}{r_0^2}-\frac{1}{r_1^2}\right)\frac{\ln \frac{r}{r_0}}{\ln \frac{r_1}{r_0}}\right)
    \nonumber
    \\
    &\phantom{\frac{m\nu}{nq^2\kappa}} + \frac{\omega_c^2}{4\nu^2} \left(\frac{1}{4} \left[-(r^2 - r_0^2) + (r_1^2 - r_0^2)\frac{\ln \frac{r}{r_0}}{\ln \frac{r_1}{r_0}}\right] - 4 c_2 \ln \frac{r}{r_0} \ln\frac{r}{r_1} + 4c_2^2\left[\frac{1}{r_0^2}-\frac{1}{r^2} - \left(\frac{1}{r_0^2}-\frac{1}{r_1^2} \right) \frac{\ln \frac{r}{r_0}}{\ln \frac{r_1}{r_0}}\right] \right)\Bigg]
    \\
    R =& \frac{m\nu}{nq^2}\left[\frac{1}{\pi}\left(\frac{1}{r_0^2}- \frac{1}{r_1^2}\right) + \frac{\omega_c^2}{\nu^2} \frac{1}{16\pi} \left(r_1^2 - r_0^2 +8c_2 \ln \frac{r_1}{r_0}\right)\right]
\end{align}
where $c_1 = \frac{1}{2}\frac{r_1^2 \ln r_1 - r_0^2 \ln r_0}{r_1^2 - r_0^2}$ and $c_2 =-\frac{1}{2}\frac{r_0^2 r_1^2}{r_1^2 - r_0^2} \ln \frac{r_1}{r_0}$. Furthermore, the function $f$ is given by
\begin{align}
    f =& \frac{12}{\ln s} \frac{a_0 + \left(\frac{\omega_c r_0^2}{\nu}\right)^2 a_2 + \left(\frac{\omega_c r_0^2}{\nu}\right)^4 a_4}{\left(R_0 + \left(\frac{\omega_c r_0^2}{\nu}\right)^2 R_2\right)^2}
\end{align}
where
\begin{align}
    a_0 =& \frac{-(s^2-1)^2 + 2(s^4-1)\ln s}{4\pi^2 s^4 \ln s} 
    \\
    a_2 =& \frac{3(s^2-1)^3 - 4 s^2(s^2-1)\ln^2 s -8 s^2 (s^2+1) \ln^3 s}{64 \pi^2 s^2 (s^2-1) \ln s}
    \\
    a_4 =& \frac{-3+3s^4 +76 s^2 \ln s}{2048\pi^2} - \frac{(s^2-1)^2}{512\pi^2 \ln s} - \frac{5s^2(s^2 +1)\ln^2 s}{256\pi^2 (s^2-1)} + \frac{s^4\ln^3 s(-3+\ln^2 s)}{96\pi^2 (s^2-1)^2} - \frac{s^4(s^2 +1) \ln^4 s}{32\pi^2(s^2-1)^3}
    \\
    R_0 =& \frac{1}{\pi} \left(1 - \frac{1}{s^2}\right)
    \\
    R_2 =& \frac{1}{16\pi} \left(s^2 - 1 - \frac{4s^2 \ln^2 s}{s^2 -1} \right)
\end{align}
and $s \equiv \frac{r_1}{r_0}$ is the aspect ratio and $R = \frac{m\nu}{nq^2 r_0^2}\left(R_0 + \left(\frac{\omega_c r_0^2}{\nu}\right)^2 R_2 \right)$. 

\begin{figure}
    \centering
    \includegraphics[width=.5\columnwidth]{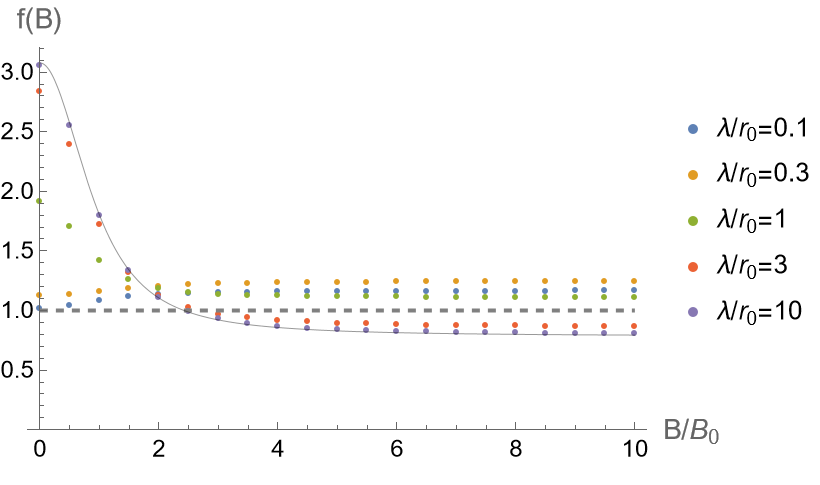}
    \caption{A plot of $f(B)$ for $s=4.5$ in the Corbino geometry. The dashed line corresponds to the ohmic $\lambda = 0$ value, while the solid thin line corresponds to the viscous $\lambda \rightarrow \infty$ value. These curves are plotted against $B/B_0$, where the scale $B_0 = (m/q) (\gamma + \nu/r_0^2)$ is different for each curve. }
    \label{fig: f(B) unnormalized}
\end{figure}

In Fig.~\ref{fig: f(B) unnormalized}, we plot $f(B)$ as a function of magnetic field for $s = 5$. At $B=0$, we see that $f(0)$ is significantly enhanced for large $\lambda/r_0$. The value of this peak depends on the aspect ratio $s$ and diverges in the $s \rightarrow 1$ limit. This ultimately results in a finite value for $T_\text{JN}$ since this divergence is compensated by the fact that $R_\text{th} \rightarrow 0$. This is strikingly different from the ohmic case, where $T_\text{JN} \rightarrow 0$ since $R_\text{th} \rightarrow 0$ as $s\rightarrow 1$.

\end{document}